\documentstyle[editedvolume,numreferences]{crckapb10} 
 
\def\gs{\mathrel{\raise1.16pt\hbox{$>$}\kern-7.0pt
\lower3.06pt\hbox{{$\scriptstyle \sim$}}}}
\def\ls{\mathrel{\raise1.16pt\hbox{$<$}\kern-7.0pt
\lower3.06pt\hbox{{$\scriptstyle \sim$}}}}

\newcommand{\stt}{\small\tt}
\begin{opening}
\title{Radio-optical orientation of E/S0 galaxies\,:}
\subtitle{APM versus FIRST}
\author{E. A. STENGLER-LARREA}
\institute{Inst.\,de F\'{\i}sica\,de\,Cantabria, Av.\,los Castros, 39005 Santander, Spain}
\author{H. ANDERNACH}
\institute{Dpto.\,de Astronom\'{\i}a, IFUG, Apdo.\,Postal 144, Guanajuato, M\'exico}

\end{opening}
\runningtitle{RADIO-OPTICAL ALIGNMENTS}
\begin{document}
\ifx\href\undefined\else\errmessage{Don't use hypertex!}\fi
\vspace*{-7.5cm}
\begin{footnotesize}\baselineskip 10 pt\noindent
Proc. {\it Observational Cosmology with the new Radio Surveys}, Tenerife,
Spain, Jan.\,13--15, 1997 \\
eds.~~M.\,Bremer, N.\,Jackson \& I.\,P\'erez-Fournon, Kluwer Acad.\,Press,
in press
\end{footnotesize}
\vspace*{6.4cm}

\begin{abstract}
We searched for extended radio sources in isolated E/S0 galaxies comparing the 
FIRST and APM catalogues for a single POSS plate. The 35 most promising 
candidates were visually inspected on the Digitized Sky Survey (DSS) and 
on FIRST images: we find several spirals and interacting galaxies and a few
E/S0s with very weak, marginally extended radio cores. The only double-lobed
(previously known) radio source is a dumbbell. For the rest of the objects,
all hosting small and weak radio sources, the DSS is inadequate 
to determine morphological types.
Thus a significant increase in sample size will be a major effort. 
\end{abstract}

%\section{Introduction}

Various studies of low-redshift, radio-emitting, early-type galaxies found
that the major axis of the radio source shows only a weak preference, if any,
to be oriented perpendicular to the optical major axis
(e.g.\,\cite{And95}). This is very much in contrast to the powerful
high-redshift radio galaxies which tend to have their radio and optical
major axes aligned \cite{Cham93}. However, the finding \cite{And93} 
that the radio-optical difference angles of brightest ellipticals in rich, 
low-z clusters is {\it bimodal} (with a narrow peak at 0$^{\circ}$ 
and a broad one at 90$^{\circ}$),
seems to support a possible evolutionary link between powerful high-z 
galaxies and dominant galaxies in clusters.

As the size of the low-z samples is still limited, we searched for 
more isolated, unperturbed, E/S0 galaxies with extended radio sources, 
comparing the 96\,May\,28 version of the FIRST radio catalogue \cite{FIRST} 
with the APM catalogue \cite{APM} of optical objects detected on the Palomar
Sky Survey.  To allow a safe determination of both
morphological type and optical major axis of the galaxies, we limit
the APM objects to m$_{\rm APM}<$17 mag in both R and O.
We consider all radio sources with deconvolved size $>$3$''$,
including those flagged as possible sidelobes, as they will hardly
coincide with a bright E/S0 galaxy by chance.

To allow for those extended radio sources appearing as two or more 
nearby components in the FIRST catalogue we also included 
multiple sources, i.e. those located within a circle of 70$''$.
APM counterparts were searched within 10$''$ of isolated radio 
sources and within 20$''$ of each of the components of a multiplet.
We also searched a radius of 20$''$ around the geometric centre
of doubles to include any optical object up to 10$''$ from the line 
joining a radio pair up to 70$''$ wide.

To test our selection criteria to find objects suitable to study the 
radio-optical orientation of E/S0s, we compared APM and FIRST~ for
POSS plate E/O\,1342, covering $\alpha$,$\delta$(J2000)=
[8:45...9:15; +32$^{\circ}$14$'$...+38$^{\circ}$25$'$].
Including both extended and multiple radio sources, we found
35 APM objects fulfilling the above criteria. 

Knowing the problems in fitting complex multiple objects in both 
the APM and FIRST catalogues we visually inspected the DSS and FIRST 
images. On DSS we found 5 multiple objects (blended into one extended 
APM object), 7 spiral-like objects, 2 interacting ones, and
3 too faint to classify or disturbed. For the remaining 18 objects,
seemingly adequate for our purposes judging from DSS, we searched
NED. This revealed 3 spirals and 7 IR-emitting objects.
The latter are most likely to be of starburst or late type. This
confirms that a morphological classification on DSS for m$\gs$15 objects
is mere guesswork and that POSS {\it plates} or {\it prints} are required 
to derive the optical type more reliably.  DSS2 scans happened to be 
available for this region, but proved to be equally insufficient for this task.

As a by-product we 'rediscovered' two of Arp's peculiar galaxies. 
For Arp\,195, a chain of 3 galaxies, FIRST shows the central and southern 
one to be radio emitters.
Arp\,202 (NGC\,2179) shows very diffuse radio emission, apparently
not reported before.
The only well-extended, double-lobed radio galaxy we found was 
B2\,0908$+$37, but as it is identified with a dumbbell galaxy \cite{Par91},
it is equally useless for our original aim.
Only for three of the E/S0s we found redshifts. Their radio power 
is near log\,P$_{\rm 1.4}$(W/Hz)=21.5, i.e.~similar to E/S0s discussed
in \cite{Sad89}.

%In conclusion, we are left with about three E/S0-like objects, although 
Our comparison of APM and FIRST over 1 POSS plate yielded about three candidate
radio E/S0s, albeit with small and faint radio sources. Not having 
inspected plates or prints, it is difficult to extrapolate our eventual success rate. 
Anticipating 3 such objects per POSS plate we may expect to find
extended radio sources in up to $\sim$500 E/S0s when FIRST will have 
reached its design goal of mapping 10$^4$ sq.\,degrees.
With only a single object per plate and no continuation of FIRST we may still
expect up to $\sim$100 objects, adequate to improve studies of the optical 
axiality of E/S0s from their radio axes \cite{San87}. Our method also
appears well-suited to extend the radio luminosity function
of E/S0s to lower luminosities and to find many more radio-weak, 
IR-emitting late type galaxies. The latter will be useful to study 
e.g.~the faint end of the radio-IR correlation. \\[-2.ex]

We thank R.L.\,White for the FIRST map server, and the {\stt SkyView} 
and NED teams for their service. H.A.~obtained a travel grant from the
conference organizers. \\[-4.ex]

{}

\end{document}